\documentclass[twocolumn,showpacs,preprintnumbers,amsmath,amssymb,prl]{revtex4}
\usepackage{graphicx}% Include figure files
\usepackage{dcolumn}% Align table columns on decimal point
\usepackage{bm}% bold math
\usepackage{epsf,epsfig,latexsym}

\begin{document}
\title{Has Nucleonic Matter at the Critical Point Been Produced in Recent 
Multifragmentation  Experiments?}

\author{J.B. Natowitz}
\author{K. Hagel}
\author{Y. Ma}
\author{M. Murray}
\author{L. Qin}
\author{S. Shlomo}
\author{R. Wada}
\author{J. Wang}

\affiliation{Cyclotron Institute, Texas A\&M University, College Station, 
Texas 77843}

\date{\today}

\begin{abstract}
For nuclei in five different mass regions the critical temperatures, at which 
the surface tension vanishes, are derived from information on the liquid 
branch of the coexistence curve. These critical temperatures increase with 
increasing mass and, for higher masses are well above recently reported 
critical temperatures obtained from Fisher Droplet model and percolation 
model analyses. However, for the lowest mass region nuclei, with 
$30<{\rm{}A}<60$, this analysis indicates that nucleonic matter has been 
produced essentially at the critical point, characterized by T$_{\rm{}c}$ and 
$\rho_{\rm{}c}$.
\end{abstract}

\maketitle

The experimental and theoretical study of possible critical behavior in a 
nuclear fluid has occupied a considerable place in many contemporary 
studies\cite{chomaz02,dasgupta01,bonasera00,moretto02,elliott02_1,kleine02,elliott02_2,raduta01}. Given the complexity\textbf{ }of nuclear 
collisions\textbf{ }there is no a priori reason why collisional excitation 
of a nucleus, with its associated dynamic evolution should lead to a 
trajectory through the critical point of the nucleonic matter phase 
diagram and ascertaining whether the actual critical point has been 
reached can be very difficult. Nevertheless, several papers have presented 
evidence for the observation of critical behavior in data from reactions 
of  1 GeV/nucleon  $^{\rm{}197}$Au, $^{\rm{}139}$La and $^{\rm{}86}$Kr  
with $^{\rm{}12}$C, taken by the EOS group\cite{hauger98,hauger00} and  
data from reactions of  high energy pions\cite{beaulieu00} and 
protons\cite{beaulieu01} with$^{\rm{}197}$Au taken by the ISiS group.  
A Fisher Droplet model analysis\cite{fisher67,fisher69} of multifragmentation 
data for  8 GeV/c $\pi^{\rm{}+}$ + $^{\rm{}197}$Au resulted in a critical 
temperature of $6.7 \pm 0.2$ MeV and 
$\rho_{\rm{}c} \approx 0.3\rho_{\rm{}0 }$\cite{elliott02_1}.  In a 
percolation analysis of 10.2 GeV/c p + $^{\rm{}197}$Au data critical 
temperature of $8.3 \pm 0.2$  MeV was determined\cite{kleine02}.  In a very 
recent paper\cite{elliott02_2}, results from a Fisher Droplet model analysis 
of the nuclear ``vapor'' have been combined with a very interesting analysis 
based upon the Principle of Corresponding States, first elucidated for 
behavior of various atomic and molecular gases\cite{guggenheim45,guggenheim93}.
In that work critical densities, 
$\rho_{\rm{}c}$ = $(0.39 \pm 0.01)\rho_{\rm{}0 }$ and critical temperatures 
of $7.6 \pm 0.2$, $7.8 \pm 0.2$, and $8.1 \pm 0.2$ MeV have been reported, 
respectively, for products of the reactions of $^{\rm{}12}$C with 
$^{\rm{}197}$Au, $^{\rm{}139}$La and $^{\rm{}86}$Kr nuclei\cite{elliott02_2}. 

In this letter we apply the corresponding states analysis to information 
derived from the liquid branch of the coexistence curve. We conclude that 
the critical temperatures of nuclei increase significantly with increasing 
nuclear mass. The values we derive are generally well above the reported 
critical temperatures, except in the lowest mass range analyzed.  In that 
mass range, $30<{\rm{}A}<60$, it appears that nuclei have been produced at 
densities and temperatures which are at, or very close to, those of the 
critical point.

We have recently carried out analyses of existing caloric curve data obtained 
in a variety of nuclear reaction studies\cite{natowitz02}. These analyses 
provided evidence for mass dependent limiting temperatures and excitation 
energies for nuclei. These limiting temperatures, and excitation energies 
are presented in Table 1.  The temperature values were found to be in good 
agreement with predicted Coulomb instability 
temperatures\cite{bonche86,levit85,besprovany89} calculated with theoretical 
models employing modern microscopic nucleon nucleon 
interactions\cite{zhang96,zhang99}. This general agreement has been exploited 
to derive the critical temperature of symmetric nuclear matter, 
T$_{\rm{}c} =16.6 \pm 0.86$ MeV  as   well as  other information on the 
nuclear equation of state\cite{natowitz02_1}.

\begin{table}[b]
\caption{Derived Temperatures and Excitation Energies}
\begin{center}
\begin{tabular}{|lllll|}
\hline
\multicolumn{5}{|c|}{Limiting and Critical Parameters (This work)} \\
\hline
Mass  & T$_{lim}$\cite{natowitz02}  & T$^{nucleonic}_{critical}$ & 
$\rho/\rho_0$ & E$^*$ \\ 
\hline
45  & $9.00 \pm 1.17$ & $8.52 \pm 0.11$ & $0.41 \pm 0.004$ & $7.71 \pm 1.0$ \\
80  & $6.97 \pm 0.60$ & $8.67 \pm 0.92$ & $0.42 \pm 0.016$ & $4.26 \pm 0.5$ \\
120 & $6.28 \pm 0.50$ & $10.1 \pm 0.87$ & $0.42 \pm 0.009$ & $3.58 \pm 1.0$ \\
160 & $6.63 \pm 0.71$ & $12.4 \pm 0.99$ & $0.41 \pm 0.005$ & $3.99 \pm 0.5$ \\
210 & $5.84 \pm 0.41$ & $12.8 \pm 2.99$ & $0.42 \pm 0.015$ & $2.53 \pm 0.3$ \\
\hline
\multicolumn{5}{|c|}{Critical Parameters (Earlier work)} \\
\hline
Mass  & T$_{crit}$  & Reference & $\rho_c/\rho_0$ & E$^*_{crit}$ \\ \hline
65 & $8.1 \pm 0.2$ &\onlinecite{elliott02_2} & $0.39 \pm 0.01$ & $4.6 \pm 0.2$
\\	
106& $7.8 \pm 0.2$ &\onlinecite{elliott02_2} & $0.39 \pm 0.01$ & $4.9 \pm 0.2$
\\
140& $7.6 \pm 0.2$ &\onlinecite{elliott02_2} & $0.39 \pm 0.01$ & $5.1 \pm 0.2$
\\
160& $6.7 \pm 0.2$ &\onlinecite{elliott02_1} & $0.3    $       & $3.8 \pm 0.3$
\\
168& $8.3 \pm 0.2$ &\onlinecite{kleine02}    & $-      $       & $3.45       $
\\
\hline
\end{tabular}
\end{center}
\end{table}

The critical temperatures and excitation energies derived in 
references\cite{elliott02_1,kleine02,elliott02_2} are also listed in Table 1 
together with the masses of the disassembling nuclei actually sampled at 
those critical excitation 
energies\cite{hauger98,hauger00,beaulieu00,beaulieu01}. They are seen to be 
reasonably close to the limiting temperatures derived from the caloric curve 
measurements in the same mass range\cite{natowitz02}. Limiting temperatures 
resulting from Coulomb instabilities would normally be expected to fall below 
the critical temperatures of the corresponding nuclei as can be easily 
demonstrated in theoretical calculations by turning off the Coulomb 
interaction\cite{bonche86}. Thus the closeness of the temperatures at which 
apparent critical behavior is observed to the limiting temperatures seen in 
the caloric curves is somewhat surprising. In assessing the significance of 
the differences seen it should be noted that the excitation energies and 
the temperatures derived from the caloric curve analyses are those 
corresponding to the onset of the entry into a caloric curve 
plateau\cite{natowitz02} while the excitation energies extracted from the 
percolation and Fisher droplet model analyses are those corresponding to 
the points where apparent critical behavior is 
observed\cite{elliott02_1,kleine02,elliott02_2}. In those works the 
temperatures are then derived from the excitation energies assuming a 
Fermi gas behavior and assigning a level density parameter. As a result 
these latter temperatures are not necessarily the same as the experimental 
values reported at those excitation energies. This by itself may be 
responsible for some of the temperature differences observed in Table 1 and 
suggests that a comparison of the excitation energies is perhaps more 
meaningful. As seen in the table the critical excitation energies tend to be 
slightly higher than the excitation energies at the entrance into the plateau.

In reference\cite{natowitz02_2} it was pointed out that an analysis of the 
caloric curves, carried out assuming a nondissipative uniform Fermi gas 
model, indicates a rapidly increasing expansion of the nuclei with 
increasing excitation energy above the excitation energy where the limiting 
temperatures are reached. In that work an iterative technique was employed 
to derive self consistent values of the thermal excitation energy, 
$\epsilon_{th}$, and the relative density, $\rho_{eq}/\rho_0$.  The 
densities derived in the early part of the plateau region of 
the caloric curve region were found to be 
$\approx 0.60 - 0.75\rho_{\rm{}0}$\cite{natowitz02_2}. Such densities   
are generally well above the values of 0.3 $\rho_{\rm{}0 }$to 
0.39$\rho_{\rm{}0 }$ derived at similar excitation energies from the Fisher 
Droplet and Corresponding States analyses\cite{elliott02_1,elliott02_2}. 
Only as the excitation energy is further increased are such densities 
reached in the caloric curve analysis. (See Table 1.) Thus it appears that 
the results from the two different methods are in contradiction.

To further explore this contradiction we have attempted to carry out a 
corresponding states analysis requiring that the results be consistent 
with our knowledge of temperature and density information for the liquid 
branch of the coexistence curve. The temperatures are readily available 
from the caloric curve but, for excitation energies below the plateau 
region of the caloric curve, the technique we have employed to derive 
densities is not applicable as shell, surface and effective mass effects 
reduce the inverse level density parameter well below its Fermi gas 
value\cite{shlomo90,shlomo91}. However, as pointed out in 
reference\cite{natowitz02_2} densities estimated from a model which assumes 
a trapezoidal density distribution and phenomenological dependencies on 
temperature derived from previous works are in good agreement with results 
of microscopic model calculations. Thus in the following, we estimate the 
average densities at excitation energies below the onset of the plateau 
region using this model.  We then take these average densities, together 
with the experimental temperatures determined at those excitation energies 
as defining the liquid branch of the liquid vapor coexistence curve.

As pointed out in references \onlinecite{moretto02,elliott02_1,elliott02_2}, 
the Principle of Corresponding States as developed by Guggenheim for inert 
gases results in a universal empirical scaling of 
$\rho_{\rm{}l,v }$/$\rho_{\rm{}c}$  vs. T/T$_{\rm{}c}$ of the form.  

\begin{eqnarray}
\rho_{\rm{}l,v}/\rho_{\rm{}c} = 
1+b_1(1 - T/T_{\rm{}c})\pm b_2(1 -T/T_{\rm{}c})^{\beta}
\end{eqnarray}

Where b$_1$ and b$_2$= (1+ b$_1$) are empirical constants, $\beta$ is a 
constant equal to 0.33, the + sign between the second and third terms is 
valid for the liquid(l) branch and the -- sign is valid for the vapor(v) 
branch of the coexistence curve. In the original scaling proposed by 
Guggenheim for macroscopic systems b$_{\rm{}1 }$is $\approx 0.75$. 
Elliott {\it et al.} have pointed out that the constant appropriate to a 
microscopic system is less as is demonstrated by  with  Ising model 
calculations for different sized cubic lattices\cite{elliott02_2}. The 
analysis of Elliott {\it et al.} indicates a value of b$_1$ near 
0.3  for the nuclei studied\cite{elliott02_2}. In the following we have 
derived values of T$_{\rm{}c}$ with fixed values of  b$_1 = 0.3$ 
and b$_2= 1.3$ and  $\beta = 0.333$. We have also evaluated the 
uncertainties on the derived values of T$_{\rm{}c}$ and $\rho_{\rm{}c}$ 
which results from varying these parameters slightly.  

\begin{figure}[t]
\epsfig{file=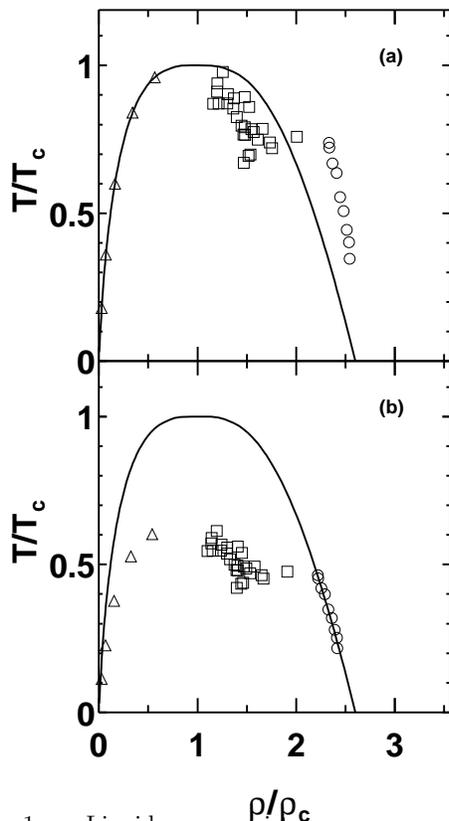,width=9.2cm,angle=0}
\vskip -1.0truecm
\caption{
Liquid gas coexistence curves. The solid line in each part of the figure 
depicts the liquid-gas coexistence curve determined in reference 
\onlinecite{elliott02_2}. Symbols 
indicate data points from the vapor  (triangles), liquid (circles) and mixed 
phase ( squares) regions. The vapor points are representative data points 
from reference \onlinecite{elliott02_2}. Liquid and mixed phase points are for 
nuclei with $140<{\rm{}A}<180$. See text. In part (a) the data are normalized 
to T$_{\rm{}c}$ = 7.6 MeV and $\rho_{\rm{}c}$ = 0.39$\rho_{\rm{}0}$. 
In part (b) the critical temperature and critical density have been 
adjusted to fit the liquid branch of the curve. There, T$_{\rm{}c}$ = 12.4 MeV
and $\rho_{\rm{}c}$ = 0.41$\rho_{\rm{}0}$.
}
\end{figure}

To illustrate the procedure which we have followed, we present in Figure 1a 
the results obtained by Elliott {\it et al.} from the analysis of the vapor 
branch obtained in 1 GeV/nucleon heavy ion reactions\cite{elliott02_2}. This 
coexistence curve is constructed from the vapor branch assuming the critical 
temperature, T$_{\rm{}c}$, is equal to 7.6 MeV\cite{elliott02_2}. The vapor 
data (which are represented by triangles) shown in the figure are 
representative points taken from reference \onlinecite{elliott02_2}. These 
points fall exactly on the coexistence curve as they are the points from 
which the curve was derived. Also shown in the figure are the data points 
derived for the liquid branch in the in the $140<{\rm{}A}<180$ mass region, 
using the approach described above (represented by circles) and the points in 
the caloric curve plateau, which are derived using the expanding Fermi gas 
hypothesis (represented by squares)\cite{natowitz02_2}. It is immediately 
apparent that the liquid branch points do not fall on the coexistence curve. 
To fit such points a significant change in T$_{\rm{}c }$, $\rho_{\rm{}c}$, 
or both, is required. In Figure 1b we present the result of fitting our 
liquid branch points of the coexistence curve to equation 1. This fit leads 
to a critical temperature of $12.4 \pm 0.99$ MeV and a critical density 
of $(0.41 \pm .005)\rho_{\rm{}0}$. Thus the liquid branch implies a much 
higher critical temperature and the overall picture which emerges from such 
a fit indicates that the limit of the Coulomb instability prevents the 
system from reaching the critical point. Disassembly in the plateau region 
of the caloric curve would then appear to be occurring in the coexistence 
region well below the critical point.

\begin{figure}[t]
\epsfig{file=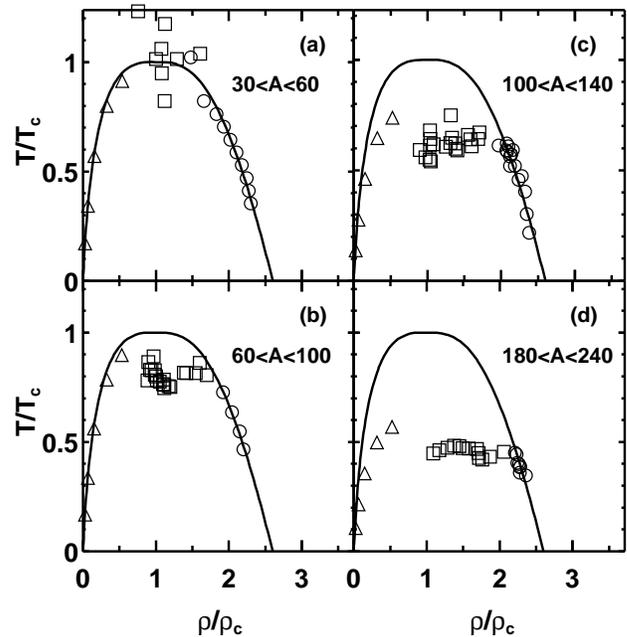,height=16cm,angle=0}
\vskip -7.0truecm
\caption{
Liquid gas coexistence curves for nuclei in four additional mass regions. The 
solid line in each part of the figure depicts the liquidgas coexistence curve 
determined in reference \onlinecite{elliott02_2}. Symbols indicate data points 
from the vapor (triangles), liquid (circles) and mixed phase (squares) 
regions. See text. In each part of the figure the critical temperature and 
critical density have been adjusted to fit the liquid branch of the curve. 
See Table 1.The Mass regions represented are (a) $30<{\rm{}A}<60$, 
(b) $60<{\rm{}A}<100$, (c) $100<{\rm{}A}<140$ and (d) $180<{\rm{}A}<240$.
}
\end{figure}

The results of extending this same method of analysis to data from the 
other mass regions are shown in Figure 2(a)-(d). The temperatures and 
densities extracted from the fits to the liquid branches are presented in 
Table 1. While the critical densities remain near 0.4$\rho_{\rm{}0, }$ the 
critical temperatures increase with increasing mass. For all but the 
lightest systems ($30<{\rm{}A}<60$) these critical temperatures are 
increasingly above those reported from the vapor branch analysis and well 
above the limiting temperatures derived from the caloric curves. 

However, for the lightest mass systems this analysis leads to a critical 
temperature of $8.5 \pm 0.11$ MeV which agrees quite reasonably with the 
$8.1 \pm 0.2$ MeV critical temperature extracted by Elliott {\it et al.} 
for their lightest target, $^{\rm{}86}$Kr\cite{elliott02_2}. Here it should 
be recalled that the masses of the nuclei studied in that technique range 
from 80 to 30 as the excitation energy deposition 
increases\cite{hauger98,hauger00}. The derived 
T$_{\rm{}c}$ for $30<{\rm{}A}<60$ is also consistent with the limiting 
temperature of $9.0 \pm 1.2$ MeV obtained from the caloric curve analysis 
in this mass region\cite{natowitz02}. \textit{Thus Figure 2(a)  provides a 
strong indication that the critical point may well have been reached in 
these light nuclei where the Coulomb effects are less important.}  For the 
heavier systems the present results suggest that the apparent critical 
behavior observed actually occurs far from the critical point. This is 
consistent with the onset of important Coulomb 
instabilities\cite{raduta01,bonche86,levit85,besprovany89,zhang96,zhang99}.  
Recent theoretical investigations have indicated that finite size effects 
or growth of fluctuations in the spinodal region may mimic critical 
behavior\cite{chomaz99,norenberg02}.

\begin{figure}[t]
\epsfig{file=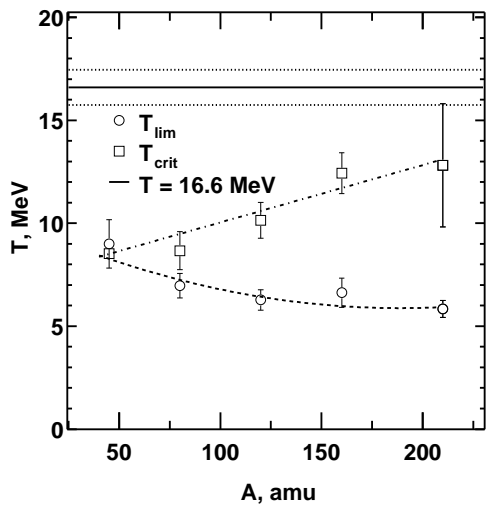,width=9.1cm,angle=0}
\vskip -5.0truecm
\caption{
Mass dependence of limiting and critical temperatures. The measured limiting 
temperatures (circles) and derived critical temperatures(squares) for nuclei 
are plotted against average mass for the different mass windows analyzed. The 
dot-dashed lines through the squares (circles) is a linear fit (2nd order 
polynomial fit)  meant to guide the eye. 
Also indicated are the symmetric nuclear matter critical temperature 
derived from these limiting temperatures (solid horizontal 
line)\cite{natowitz02_1} and the uncertainty associated with that 
value (dotted horizontal lines).
}
\end{figure}

The results of this analysis are summarized in Figure 3. There the critical 
temperatures derived from fits to the liquid branch are compared with the 
limiting Coulomb instability temperatures. Also indicated by a horizontal 
bar in the figure is the critical temperature for symmetric nuclear matter, 
previously derived from the limiting temperature data.
While the limiting temperatures drop with increasing mass, reflecting the 
increased importance of Coulomb effects, the critical temperatures rise 
toward the infinite nuclear matter value. The higher values for heavier 
nuclei are in accord with the constraint that T$_{\rm{}c}$ should be 
greater than 10 MeV as suggested by Karnaukhov based upon an analysis 
of fission probabilities taking into account the temperature dependence 
of the surface energy and its effect on the fission barrier\cite{karnaukhov97}

%Previous calculations using Fermionic Molecular dynamics\cite{feldmeier00} 
%or a microcanonical model\cite{hauger98} have both suggested critical 
%temperatures near 8-9 MeV for light nuclei. The higher values for heavier 
%nuclei are in accord with the constraint that T$_{\rm{}c}$ should be greater 
%than 10 MeV as suggested by Karnaukhov based upon an analysis of fission 
%probabilities taking into account the temperature dependence of the surface 
%energy and its effect on the fission barrier\cite{karnaukhov97}.

\begin{flushleft}
Acknowledgements
\end{flushleft}

This work was supported by the U S Department of Energy under Grant DEFG03 
93ER40773 and by the Robert A. Welch Foundation.


\begin{thebibliography}{99}
\bibitem{chomaz02}P. Chomaz, Proceedings of the INPC 2001 Conference, 
Berkeley, Ca., July, 2001.

\bibitem{dasgupta01}S. Das Gupta, A. Z. Mekjian, M. B. Tsang, nucl-th/0009033, 
LANL preprint server, submitted to Adv. Nucl. Phys.(2001).

\bibitem{bonasera00}A. Bonasera, M. Bruno, C.O. Dorso and P.F. Mastinu, 
Riv. Del Nuov. Cim. {\bf 23}, 1 (2000). 

\bibitem{moretto02}L.G. Moretto, J. B. Elliott, L. Phair, G. J. Wozniak, 
C. M. Mader and A. Chappars, Preprint LBNL-49235, to appear in Proceedings 
of the INPC 2001 Conference, Berkeley, Ca., July, 2001.

\bibitem{elliott02_1}J. B. Elliott, L. G. Moretto, L. Phair, G. J. Wozniak, 
T. Lefort, L. Beaulieu, K. Kwiatkowski, W.C. Hsi, L. Pienkowski, H. Breuer, 
R. G. Korteling, R. Laforest, E. Martin, E. Ramakrishnan, D. Rowland, 
A. Ruangma, V. E. Viola, E. Winchester, S. J. Yennello, Phys. Rev. Lett. 
\textbf{88}, 042701 (2002).

\bibitem{kleine02}M. Kleine Berkenbusch {\it et al.}, Phys. Rev. Lett. \textbf{88}, 
022701 (2002).

\bibitem{elliott02_2}J. B. Elliott {\it et al.}, LBNL-29237, submitted to 
Phys. Rev. C (2002). ArXiv nucl-ex/0205004.

\bibitem{raduta01}H. Raduta and A. Raduta, Phys. Rev. Lett. {\bf 87}, 202701 
(2001).

\bibitem{hauger98}J. A. Hauger {\it et al.}, Phys. Rev. C {\bf 57}, 764 (1998).

\bibitem{hauger00}J. A. Hauger {\it et al.}, Phys. Rev. C {\bf 62}, 024626 
(2000).

\bibitem{beaulieu00}L. Beaulieu \textit{et al.}, Phys. Rev. Lett. \textbf{84},
5971 (2000).  

\bibitem{beaulieu01}L. Beaulieu \textit{et al.}, Phys. Rev. C \textbf{63}, 
031302 (2001). 

\bibitem{fisher67}M. E. Fisher, Physics {\bf 3}, 255 (1967).

\bibitem{fisher69}M. E. Fisher, Rep. Prog. Phys. {\bf 30}, 615 (1969).

\bibitem{guggenheim45}E. A. Guggenheim, J. Chem. Phys., {\bf 13}, 253 (1945).

\bibitem{guggenheim93}E.A. Guggenheim, ``Thermodynamics",
4th ed. (NorthHolland, 1993).

\bibitem{natowitz02}J.B. Natowitz, R. Wada, K. Hagel, T Keutgen, M. Murray,
Y. G. Ma , A. Makeev, L. Qin , P. Smith and C.Hamilton, Phys. Rev. C {\bf 65},
034618 (2002).

\bibitem{bonche86}P. Bonche\textit{,} S. Levit and H. Vautherin, Nucl. Phys. 
A{\bf 427}, 278 (1984); Ibid, A {\bf 436}, 265 (1986)P. Bonche, S. Levit, 
and D. Vautherin, Nucl. Phys. A{\bf 436}, 265 (1985).

\bibitem{levit85}S.Levit and P.Bonche, Nucl. Phys. A{\bf 437}, 426 (1985).

\bibitem{besprovany89}J. Besprovany and S. Levitt Phys. Lett. B{\bf 217}, 1 
(1989).

\bibitem{zhang96}Y. Zhang, R. Su, H. Song, and F. Lin, Phys. Rev. C {\bf 54}, 
1137  (1996). 

\bibitem{zhang99}L. L. Zhang, H. Q. Song,  P. Wang, and R. K. Su, Phys. Rev. 
C {\bf 59}, 3292 (1999).

\bibitem{natowitz02_1}J.B. Natowitz, K. Hagel, Y. Ma, M. Murray, L. Qin, 
R. Wada and J. Wang, submitted to  Phys. Rev. Letters (2002); 
ArXiv nucl-ex/0204015.

\bibitem{natowitz02_2}J.B. Natowitz, K. Hagel, Y. Ma, M. Murray, L. Qin, 
S. Shlomo, R. Wada and J. Wang, submitted to  Phys. Rev. C (2002); 
ArXiv nucl-ex/0205005.

\bibitem{shlomo90}S. Shlomo and J.B. Natowitz, Phys. Lett. B {\bf 252}, 187 
(1990).

\bibitem{shlomo91}S. Shlomo and J.B. Natowitz, Phys. Rev. C \textbf{44}, 
2878 (1991).

\bibitem{chomaz99}P. Chomaz and F. Gulminelli, 
Phys. Lett. B {\bf 447}, 221 (1999).

\bibitem{norenberg02}W. Norenberg, G. Papp and P. Rozmej GSI Preprint 20023, 
January, 2002.

\bibitem{karnaukhov97}V.A. Karnaukhov, Phys. Atom. Nucl. {\bf 10}, 1625 (1997).

%\bibitem{feldmeier00}H. Feldmeier and J. Schnack Rev. Mod. Phys. 72, 655 
%(2000).

%\bibitem{karnaukhov_2}Karnaukhov

%\bibitem{kertesz}Kertesz

%\bibitem{sugawa99}Y. Sugawa and H. Horiuchi, Phys. Rev. C 60, 607 (1999).

%\bibitem{chomaz_2}Chomaz has suggested that the point of Coulomb instability 
%may be viewed as the point at which the nucleus 
%e\textbf{nters into the spinodal region???.In the spinodal region the
%growth of fluctuations  Norrenberg} Further, finite size effects Kertesz line

%\bibitem{chomaz_3}\textbf{
%\textit{Critical point }}\textit{( Sobotka)}\textbf{\textit{. 
%Critical behavior, Kertesz line, Spinodal etc>}}
\end{thebibliography}
\end{document}